## Gravity can be caused by the difference of Coulomb's constants.

**Yoji Hagiya,** Kawasaki Kanagawa Japan

### Abstract

**Purpose**

The author reveals the identity of gravity by ensuring that the net coulomb force between two objects is gravity.

**Method**

Coulomb's constant is defined as a value for both attraction and repulsion. However, it is strange that only one value can be used for both attraction and repulsion.

A very small difference between Coulomb's constant for attraction and Coulomb's constant for repulsion can be the source of gravity. The author has verified that this theory is correct by calculating with a slightly smaller Coulomb's constant for repulsion.

**Result**

It is very likely that the source of gravity is the difference between the Coulomb constants for attraction and repulsion.

### Introduction

Although 300 years have passed since Isaac Newton discovered gravity, it is still not known what mechanism causes it.

Scientists believe that gravity and coulomb force are very similar, but usually they think that gravity and coulomb force are different things because of their characters.

In 1769, Scottish physicist John Robison [1] announced that, according to his measurements, the coulomb force of repulsion between two objects with charges of the same sign varied as x^-2.06. This suggests that coulomb force of attraction is stronger than coulomb force of repulsion.

Albert Einstein was trying to unify gravity and electromagnetic force in his late years but it seems that he did not know John Robison's experiment and the result.

An object contains equal number of electrons and protons. A neutron is considered as a pair of electron and proton. So, if we take two objects:

Electrons in object1 repel electrons in object2,

Electrons in object1 attract protons in object2,

Protons in object1 attract electrons in object2,

Protons in object1 repel protons in object2.

Based on superposition principle, the sum of the forces above can be regarded as net coulomb force between the two objects.

## Method

To verify that the gravity between two objects is the net coulomb force between those two objects, expand the equation of coulomb's law based on superposition principle and calculate each term independently then add those forces.

To calculate net coulomb force between two objects, following equation is used:

$$F_g = k_r \frac{q_{1e} q_{2e}}{r^2} + k_a \frac{q_{1e} q_{2p}}{r^2} + k_a \frac{q_{1p} q_{2e}}{r^2} + k_r \frac{q_{1p} q_{2p}}{r^2}$$

$F_g$ : Force of gravity
$k_r$ : Coulomb's constant for repulsion
$k_a$ : Coulomb's constant for attraction
$q_{1e}$ : Signed magnitude of electrons in object 1
$q_{1p}$ : Signed magnitude of protons in object 1
$q_{2e}$ : Signed magnitude of electrons in object 2
$q_{2p}$ : Singed magnitude of protons in object 2
$r$ : Distance between the objects

Neutron is considered as a pair of electron and proton.

As the parameters of the calculation, following data are used (Units are SI):
Coulomb's constant for:
attraction: 8.987551792e+9
repulsion: 8.987551791999999999999999999999999996e+9 (estimated value)
Mass of sun: 1.9884e+30 kg
Mass of earth: 5.9724e+24 kg
Distance between the sun and the earth: 1.495978e+11 m
G: 6.67384e-11
Charge of electron: -1.602176634e-19 C
Charge of proton: 1.602176634e-19 C
Mass of electron: 9.109383701e-31 kg
Mass of proton: 1.672621898e-27 kg

## Result

Using my estimated value of the Coulomb constant of Coulomb repulsion, I calculated the gravity between the Sun and the Earth as follows.

Force of gravity calculated based on Newton's law: 3.5414247509804215e+22 N
Net coulomb force calculated by the method above: -3.890E+22 N

Substituting a value slightly smaller than the Coulomb constant of Coulomb's attraction into the Coulomb constant of Coulomb's repulsion, the calculation was in close agreement with the gravity obtained by Newtonian mechanics.

## Discussion

It has become clear that slight difference between coulomb's constant for attraction and coulomb's constant for repulsion could cause the gravity.

Such difference is not measured yet because it is so small and very hard to measure by the current technology so far.

Usually, scientists say that neutrino respond to the gravity but it has no charge. However, because gravity is coulomb force, if it responds to the gravity, it seems to have some charge.

## Acknowledgement

The author would like to thank Dr. Tomio Petrosky, Dr. Kazuo Furukawa, Mr. Hideki Masudaya, and Mr. Masatoshi Sou for the advice.